# Escape Time of Josephson Junctions for Signal Detection

P. Addesso, G. Filatrella and V. Pierro

**Abstract** In this Chapter we investigate with the methods of signal detection the response of a Josephson junction to a perturbation to decide if the perturbation contains a coherent oscillation embedded in the background noise. When a Josephson Junction is irradiated by an external noisy source, it eventually leaves the static state and reaches a steady voltage state. The appearance of a voltage step allows to measure the time spent in the metastable state before the transition to the running state, thus defining an escape time. The distribution of the escape times depends upon the characteristics of the noise and the Josephson junction. Moreover, the properties of the distribution depends on the features of the signal (amplitude, frequency and phase), which can be therefore inferred through the appropriate signal processing methods. Signal detection with JJ is interesting for practical purposes, inasmuch as the superconductive elements can be (in principle) cooled to the absolute zero and therefore can add (in practice) as little intrinsic noise as refrigeration allows. It is relevant that the escape times bear a hallmark of

P. Addesso
Department of Electronic and Computer Engineering, University of Salerno,
Via Ponte Don Melillo, 1, 84084, Fisciano, Italy
e-mail: paddesso@unisa.it

G. Filatrella
Department of Sciences for Biological, Geological, and Environmental Studies
and Salerno unit of CNSIM, University of Sannio, Via Port'Arsa 11, 82100,
Benevento, Italy
e-mail: filatrella@unisannio.it

V. Pierro
Department of Engineering, University of Sannio, Corso Garibaldi, 107,
82100, Benevento, Italy
e-mail: pierro@unisannio.it

the noise itself. The spectrum of the fluctuations due to the intrinsic classical (owed to thermal or environmental disturbances) or quantum (due to the tunnel across the barrier) sources are different. Therefore, a careful analysis of the escape times could also assist to discriminate the nature of the noise.

# 1 Introduction

The Josephson effect, early observed in superconductivity [4], essentially consists in particles tunneling between two weakly coupled systems, each described by a macroscopic wave functions. The effect is for instance observed when two superconductors are placed close enough (few nanometers) to let the bosonic waveforms of the charge carriers (Cooper's pairs) overlap. However, The effect is generic, and has been predicted [28] and observed [39] also between Bose-Einstein bosonic condensates [15]. The main features of the Josephson effect in superconductivity are the possibility of a non dissipative flow of Cooper's pairs from one superconductor to the other (the so called d.c. Josephson effect) and the appearance of a voltage proportional to the derivative of the phase difference between the superconductors' wave functions (the a.c. Josephson effect). The two effects give rise to a nonlinear device, a Josephson Junction (JJ). A Josephson Junction (JJ) can be used as a threshold detector [13, 46], i.e. as a device capable to discriminate a signal with a sharp transition from a state to another [1, 23]. Threshold detection is a suboptimal tool in signal processing, because in typical applications the optimal choice is linear matched filtering. However in some circumstances analogue devices might prove competitive, for optimal strategies cannot be efficiently implemented, when: ($i$) the amount of data necessary to claim a detection at very low Signal to Noise Ratio (SNR) is too high (such as all-sky all-frequency search of gravitational wave emitted by a pulsar [34, 49]); ($ii$) the amplitude of the signal is too low, and amplification introduces a significant additional noise; ($iii$) the frequency band of the signal falls in the THz range, where standard digital sampling procedures fails (e.g. THz sensing [45]).

When numerical analysis of the sampled data is impractical one should resort analogue techniques. A hysteretic JJ can be employed as signal detector, because a sinusoidal waveform corrupted by noise eventually forces the junction to switch from the zero voltage to a finite voltage state. The switch occurs when the gauge invariant phase crosses a threshold: the maximum of the potential that hooks the phase in the superconducting state. The appearance of a voltage step thus flags that the JJ phase has exceeded the threshold and makes it possible to measure how much time the JJ has been in the superconducting state before to overcome the energy barrier. The measure of the escape time in JJ is a well established technique, inasmuch it has been for more than 50 years a standard tool to characterize the properties of the metastable state [3]. For instance it has been employed to prove the existence of Macroscopic Quantum Tunneling (MQT) [17], i.e. the possibility that the escape occurs even at zero temperature for the quantum nature of the Josephson phase [10].

The behaviour of the Josephson phase cannot be directly observed, it is just possible to detect the change of the voltage (proportional to the phase derivative) associated with the passage over the barrier. Moreover, it is not possible to recover the Josephson phase sampling the derivative (i.e. the junction voltage). In fact the measured signal is altered by the application of a very selective filtering stage to avoid that environmental noise reaches the device. Moreover, voltage oscillations fall on a frequency scale (close to the Josephson frequency) outside the band of conventional electronic devices. Thus it makes good-sense to assert that the available information is to be confined to the escape time.

From the point of view of data analysis, a further loss of information occurs when the analysis of the escape time sequence is limited to the average lifetime. Apart from the simplicity of the analysis, so much information can be depleted because the average escape time retains most of the physics of the escape from a static potential. In fact in Kramer theory [50] the average escape time is related to the most important parameter of the escape process, the ratio between the potential well and the intensity of the fluctuations. Recently, suitable statistics (e.g. variance and skewness) of the lifetime distribution have been analyzed to highlight the Poissonian nature of the process [48] or to detect the transition from the under-damped to moderately damped regime [20, 42]. Detection theory allows to take a further step and to exploit the full distribution of the escape times to improve the performances of the JJ as a detector [1]. In this Chapter after the discussion of the JJ model (Sect. 2), we will show how the analysis of the distribution of the escape times can be performed with the methods of signal processing (Sect. 3) The main features of detectors based on JJ are described in Sect. 4. In Sect. 5 we discuss about some practical issues in real experimental scenarios together with a brief discussion about the possibility of distinguishing between the quantum or classic nature of the noise. The last Sect. 6 is, as usual, devoted to the conclusions.

## 2 Physical Model of a JJ Detector

The dynamic variable of a JJ is the gauge invariant superconducting phase, $\varphi$, ruled by the celebrated Josephson equations [4]:

$$\begin{aligned} I &= I_c \sin(\varphi) \\ V &= \frac{\hbar}{2e} \frac{\mathrm{d}\varphi}{\mathrm{d}t} \end{aligned} \quad (1)$$

$I_c$ denotes the Josephson critical current, or the maximum current of Cooper's pairs (whose charge is $2e$) which can tunnel without an applied voltage. However, a practical device is far more complicated than this. First, a real junction is also characterized by a resistance and a capacitance, second the JJ is dc and ac biased, and third the environment at finite temperature disturbs the junction with noise. The resulting Langevin model equation of a JJ biased with a sinusoidal signal corrupted by additive noise $\xi(t)$ and subject to thermal fluctuations $n(t)$ reads [23]:

$$\frac{C\hbar}{2e}\frac{d^2\varphi}{dt^2} + \frac{\hbar}{R2e}\frac{d\varphi}{dt} + I_c \sin(\varphi) = I_b + S_0 \sin(\Omega t + \varphi_0) + \sqrt{D}\xi(t) + \sqrt{k_B T/R}\, n(t). \quad (2)$$

Here $C$ and $R$ are the capacitance and the resistance of the JJ, respectively (we consider JJ in the underdamped regime, i.e. that the capacitance is not negligible). Furthermore $I_b$ is the dc bias current, $S_0$ the amplitude of the ac term of frequency $\Omega$ and initial phase $\varphi_0$. In Eq. (2) two random terms appear: $\xi(t)$, with intensity $\sqrt{D}$, that represents an additive noise corrupting the external signal and $n(t)$, with intensity $\sqrt{k_B T/R}$, that represents thermal current ($k_B$ denotes the Boltzmann constant and $T$ the temperature). The terms $\xi(t)$ and $n(t)$ are white Gaussian noise stochastic processes, whose correlators read $<n(t)n(t')> = 2\delta(t-t')$, and $<\xi(t)\xi(t')> = 2\delta(t-t')$. For signal detectors it is important that in Eq. (2) thermal fluctuations $\sqrt{k_B T/R}$ can be neglected with respect to the signal noise intensity $\sqrt{D}$. In fact JJ can be cooled down at a temperature $T$ much below the signal noise temperature viz. $T \ll DR/k_B$, an assumption that we will retain for the remaining of the Chapter.

The low temperature condition is favored when the junction resistance is high (i.e. when dissipation is low) because Eq. (2) is based on a parallel lumped circuit model (see the inset to Fig. 1).

To assume that the signal is corrupted only by an additive term is a simplification: noise can affect the signal in several ways, for instance as a multiplicative noise [40, 41] or phase and frequency fluctuations [29]. In the following we will limit our analysis to the standard additive noise case; however we expect that the results can be indicative of the behavior also for other noise sources. For instance frequency fluctuations of the JJ driving signal can be treated, in some limits, as an additive noise [22]. A special mention is deserved to the role of quantum fluctuations. Tunneling of the macroscopic phase $\varphi$ has been demonstrated in early measurements [16, 18] as a saturation of the escape times at very low temperatures. The experiments are performed while decreasing the bath temperature to reduce the fluctuations; if one eventually observes a constant escape rate in spite of the change in temperature, it is concluded that the rate itself should be due to some other fluctuation source, such as MQT. It has been shown that quantum fluctuations contribute as an equivalent thermal source of temperature $\theta^* = e\omega_J/(\pi I_c)$ [2] where $\omega_J = [2eI_c/(\hbar C)]^{1/2}$ is the characteristic frequency (called the Josephson frequency). If only the average escape time is considered, quantum noise and stochastic effects are nearly equivalent, with the fundamental difference that quantum noise is unavoidable and operates also in the zero temperature limit. Further details about the differences between quantum and classical noise will be reported in Sect. 3.

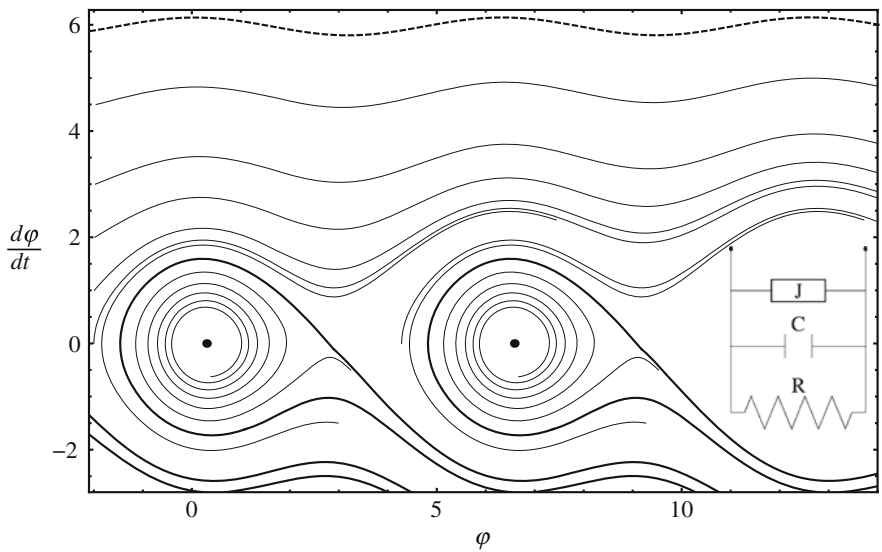

**Fig. 1** Phase plane portrait of Eq. (4) for $\gamma = 0.3$, $\alpha = 0.05$, $\varepsilon = \varepsilon_n = 0$. Thin curves denote the phase lines, while the solid thick curves denote the separatrix between the asymptotic running state (dashed curve) and the static solution (thick dots). The symbol + denotes the instable fix point. The inset shows the electric circuit model of Eq. (2)

We recast Eq. (2) introducing the dimensionless time $\tau = \omega_J t$, normalized with respect to $\omega_J$. Dividing for the critical current $I_c$, and rearranging the terms, Eq. (2) reads

$$\frac{d^2\varphi}{d\tau^2} + \frac{\omega_J}{RI_c}\frac{\hbar}{2e}\frac{d\varphi}{d\tau} + \sin(\varphi) = \frac{I_b}{I_c} + \frac{S_0}{I_c}\sin(\frac{\Omega}{\omega_J}\tau + \varphi_0) + \frac{\sqrt{\omega_J D}}{I_c}\tilde{\xi}(\tau), \quad (3)$$

where the correlator reads $<\tilde{\xi}(\tau)\tilde{\xi}(\tau')> = 2\delta(\tau - \tau')$. With the definitions $\gamma = I_b/I_c$ as the normalized bias current, $\alpha = (\omega_J/RI_c)(\hbar/2e)$ as the normalized dissipation, $\varepsilon = S_0/I_c$ as the normalized signal amplitude and $\sqrt{\varepsilon_n} = \sqrt{(\omega_J D/I_c^2)}$ as the normalized noise intensity, Eq. (3) becomes:

$$\frac{d^2\varphi}{d\tau^2} + \alpha\frac{d\varphi}{d\tau} + \sin(\varphi) = \gamma + \varepsilon\sin(\omega\tau + \varphi_0) + \sqrt{\varepsilon_n}\tilde{\xi}(\tau). \quad (4)$$

Eq. (4) is a well established framework for periodically driven stochastic systems [35]. The analysis of the asymptotic or stationary states can be performed with the analysis of the corresponding Fokker-Planck equation [50]. Unfortunately, to profitably apply the methods of signal detection it is important to accurately know the distribution of the shortest escape times, which is not easily retrieved from the analysis of the Fokker Planck equation.

The schematic of the physics of the device is depicted in Fig. 1 by the zero noise limit phase plane of Eq. (4). The thick line is the separatrix between confined oscillations and running states. In fact a JJ is also a practical realization of the prototypal washboard potential

$$U(\varphi) = -\gamma\varphi - \cos(\varphi). \tag{5}$$

For $\gamma < 1$ Eq. 4) gives rise to a barrier [4, 7]:

$$\Delta U(\gamma) = 2[\sqrt{1-\gamma^2} - \gamma\cos^{-1}(\gamma)]. \tag{6}$$

If the oscillating current is zero ($\varepsilon = 0$) for low noise ($\varepsilon_n << \Delta U$) escape occurs at a rate $r_K$ [50]

$$r_K \propto T_K^{-1} \exp\left(-\frac{\Delta U}{\varepsilon_n}\right). \tag{7}$$

Such a rate is related to the average escape time $\mu_0 = 1/r_K$ and $T_K$ is the Kramer prefactor [50]. The individual escape times can be directly measured in a variety of conditions [55, 16], including high-Tc superconductors [5, 6]; in the case of underdamped JJ ($\alpha < 1$) when the system overcomes the energy barrier $\Delta U$, it switches from the locked state to a finite voltage running state [4]:

$$\frac{<V>}{V_0} = <\frac{d\varphi}{d\tau}> \simeq \frac{\gamma}{\alpha}, \tag{8}$$

where $<\cdot>$ indicates the temporal moving average and the voltage is normalized with respect to $V_0 = \hbar\omega_J/2e$. The efficiency of this voltage switching is less evident for moderately damped (i.e. $0.25 < \alpha < 1$) JJ, when the Josephson phase after the passage over the barrier could be retrapped in another well by the noise (this corresponds in Fig. 1 to cross the separatrix in the reverse direction, from the running to the static state) [42].

The escape process is illustrated in Fig. 2 : The Josephson phase $\varphi$ fluctuates at the bottom of the washboard potential until it is driven across the separatrix (the vertical line) and the running state is reached. The elapsed time is the escape time $\tau_i$ of the metastable state.

The acquisition of escape times is a process that should be carefully performed to retain maximal information about the presence of signal. In particular, the initial phase is a crucial parameter because the signal to noise mixture can be applied to the JJ in different ways [23, 56] to acquire the escape time sequence $\underline{\tau} = \{\tau_i\}_{i=1}^N$. Indeed, after the JJ has switched to the running state there are a variety of methods to reset the system. If the frequency of the signal is perfectly known, in principle it is possible to reapply the signal with the same initial phase $\varphi_0$. This acquisition strategy is called coherent. To apply the signal again with the same initial phase some fraction of the signal is lost waiting for the correct time to restart the process (see Fig. 2).

When the frequency cannot be easily controlled or the time measurement precision is too low, another possibility arises, i.e. incoherent acquisition. It consists in reapplying the signal with any phase it might have after the reset

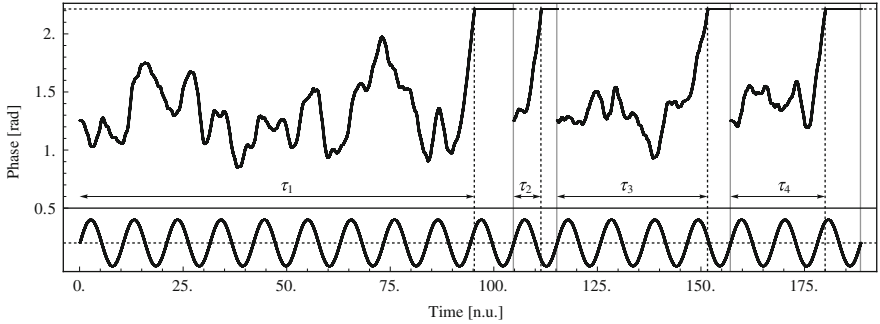

**Fig. 2** Examples of the time dependent trajectories of the phase $\varphi$ for the coherent acquisition strategy. The lower part of the figure shows the applied signal $\varepsilon \sin(\omega\tau + \varphi_0)$. The dotted vertical lines denote the transitions across the separatrix [the threshold is $\varphi = \pi - \sin^{-1}(\gamma)$] that determines the escape times $\tau_1, \tau_2, \ldots \tau_N$. After the switching has occurred, the JJ is restarted with static state initial condition ($\varphi = \sin^{-1}(\gamma)$, $d\varphi/d\tau = 0$) and the signal is applied with the same initial phase $\varphi_0$ (continuous vertical line). Parameters of the simulations are: $\gamma = 0.8$, $\alpha = 0.05$, $\varepsilon_N = 0.0175$. The signal parameters are: $\varepsilon = 0.05$, $\varphi_0 = 0$ and $\omega = 0.7$

procedure and therefore with an essentially random initial value of $\varphi_0$. We investigate two extreme experimental situations.

- Instantaneous signal reset: the initial phase coincides with the exit phase across the separatrix of the previous escape.
- Random signal reset: the initial phase is supposed uniformly distributed in $[0, 2\pi]$.

In both cases much of the information carried by the initial phase parameter is lost. In the next Section we analyze the escape time distributions and we use statistical decision theory to seek efficient methods for detection of noisy harmonic signals.

## 3 Detection Theory for Josephson Junctions Threshold Detection

Detection theory methods are aimed to discern the presence of a known waveform embedded in a random (noisy) background. In this framework the analysis of the escape times serves to discern if the time dependent bias, the right hand side of Eq. (4), consists of just the constant bias $\gamma$ and the unavoidable noise $\sqrt{\varepsilon_n}\tilde{\xi}(t)$, or contains also a sinusoidal oscillation of amplitude $\varepsilon$, frequency $\omega$ and phase $\varphi_0$.

The system is characterized by a SNR related to the ratio $\varepsilon/\sqrt{\varepsilon_n}$, which weights the signal against the intensity of the noise correlations. In the language of signal detection one formulates two hypothesis:

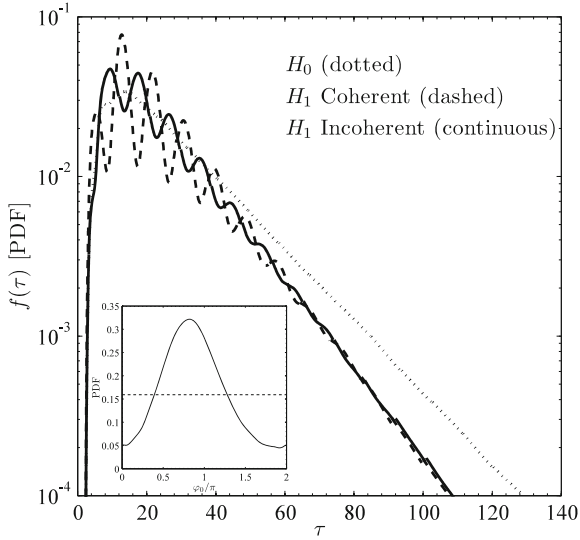

**Fig. 3** PDFs of the escape times. The three curves represent: Signal absence ($H_0$ : dotted line); signal acquired by a coherent strategy ($H_1$ : dashed line); signal acquired by an incoherent strategy with instantaneous signal reset ($H_1$ : continuous line). The inset shows the PDFs of the signal initial phase $\varphi_0$ for the incoherent acquisition with instantaneous signal reset (continuous line), and with random signal reset (dashed line). Parameters of the simulations are: $\gamma = 0.8$, $\alpha = 0.05$, $\varepsilon_N = 0.0175$. The signal parameters are: $\varepsilon = 0.05$, $\varphi_0 = 0$ and $\omega = 0.7$

- $H_0$ : the JJ is uniquely driven by noise;
- $H_1$ : the JJ is driven by noise and a sinusoidal excitation.

If one collects the escape times, obtained with numerical simulations of Eq. (4) [43], it is possible to estimate the Probability Density Functions (PDF) as in Fig. 3 . Here we show three cases: (*i*) the signal is absent ($H_0$); (*ii*) the signal is present and escape times are acquired by the coherent strategy ($H_1$ coherent); (*iii*) the signal is present and escape times are acquired by the incoherent strategy with instantaneous signal reset ($H_1$ incoherent).

It is evident that the three curves have significantly distinct shapes, even if they share an exponential decay. In particular the PDFs related to the signal presence have approximately the same slope (i.e. the same average escape time), which differs from the slope of the $H_0$ hypothesis. Moreover both $H_1$ PDFs exhibit oscillations. The oscillations, related to the signal amplitude and frequency, are most evident for the coherent acquisition strategy. In the coherent case the oscillations also depend upon the signal initial phase $\varphi_0$. Noticeably, the incoherent acquisition with instantaneous signal reset exhibits oscillations due to a synchronization of the signal exit phase around $\pi$, as can be seen in the inset to Fig. 3 (more precisely the density is peaked around $4\pi/5$). These oscillations are definitively smaller in the incoherent case with random signal reset (not shown in Fig. 3), which only preserves the same slope of the other two $H_1$ curves.

Several strategies can be designed to discriminate the presence from the absence of the sinusoidal signal. A starting point is that the PDF of the type shown in Fig. 3 are approximately exponential and exhibit a change of slope, i.e. the statistical average of escape times under the two hypotheses (say $\mu_0 = E[\tau|H_0]$ and $\mu_1 = E[\tau|H_1]$) are different. In this case, it is possible to use a detection strategy based on the Sample Mean (SM). In fact by averaging $N$ identically distributed escape times $\underline{\tau} = \{\tau_i\}_{i\in[1,N]}$, it is possible to define a detector based on the following test:

$$\mathscr{A}(\underline{\tau}) = \frac{s}{N}\sum_{i=1}^{N} \tau_i \underset{H_0}{\overset{H_1}{\gtrless}} \zeta, \qquad (9)$$

where $s = sign(\mu_1 - \mu_0)$ and $\zeta$ is a suitable threshold. The SM strategy, analyzed in detail in [23], has the great advantage of simplicity, but it is obviously suboptimal. Indeed in the coherent and the incoherent (with instantaneous signal reset) acquisition strategies escape time PDFs exhibit oscillations (see Fig. 3). Thus it is important to identify a strategy that fully exploits the information carried by the escape time statistics. the Neyman-Pearson lemma [51] indicates that an optimal strategy, based on the Likelihood Ratio Test (LRT), exists. The analysis can be performed by comparing the product of the ratio between the PDF, evaluated at the escape time samples $\tau_i$, with and without the signal. If we denote by $f_{0,1}(\cdot)$ the PDFs of the escape times under the hypothesis $H_{0,1}$, the test can be written as:

$$\mathscr{A}(\underline{\tau}) = \frac{s}{N}\sum_{i=1}^{N} \tau_i \underset{H_0}{\overset{H_1}{\gtrless}} \zeta, \qquad (10)$$

A useful transformation, by means of the log nonlinearity, leads to the statistics:

$$\Lambda(\underline{\tau}) = \frac{1}{N}\sum_{i=1}^{N} \log\left[\frac{f_1(\tau_i)}{f_0(\tau_i)}\right] \underset{H_0}{\overset{H_1}{\gtrless}} \zeta. \qquad (11)$$

where the new threshold is $\zeta = \log(\zeta')/N$.

Optimality of the LRT concerns the minimization of the error rate. Two types of error assess the quality of a detector:

- the *false alarm probability* $P_f$, also known as Type I error probability, i.e. the probability to decide for the hypothesis $H_1$ when $H_0$ is true;
- the *miss probability* $P_m$, also known as Type II error probability, i.e. the probability to decide for the hypothesis $H_0$ when $H_1$ is true.

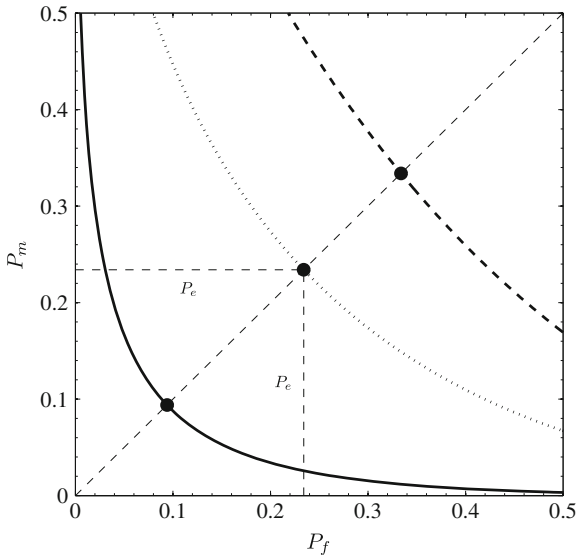

**Fig. 4** Typical ROCs of a JJ based detector under the hypothesis of complete parameter knowledge. The three curves are related to: LRT Coherent (continuous line); LRT Incoherent with instantaneous signal reset (dotted line); SM Coherent (dashed line). Parameters of the simulations are: $\gamma = 0.8$, $\alpha = 0.05$, $\varepsilon_N = 0.0175$. The signal parameter are $\varepsilon = 0.05$ and $\omega = 0.7$. The simulations are performed setting the mean observation time under $H_0$, $E[T_{obs}|H_0] = 500$

Under the hypothesis that the JJ parameters ($\alpha$ and $\gamma$), the noise variance $\varepsilon_n$ (which does not depend on the particular hypothesis in force) and the signal parameters ($\varepsilon$, $\omega$ and $\varphi_0$) are *perfectly known*, LRT minimizes, among all possible tests, the miss probability $P_m$ at a fixed false alarm level $P_f$. In this framework, both $P_f$ and $P_m$ are functions of the threshold $\zeta$. Thus a popular way to summarize the results of signal analysis is to compute the Receiver Operator Characteristic (ROC) of the test statistic, that is the plot of $P_f$ versus $P_m$ for different values of $\zeta$.

A ROC example is presented in Fig. 4, in which the unavoidable trade-off between the two error probabilities is shown. The performance of the LRT strategies, especially in the case of coherent acquisition, are significantly better than the SM ones. For sake of clarity, we have shown only one SM curve (in the coherent case), because in the other cases the performances are approximately the same. Noticeably, due to the lack of oscillations, in the incoherent random signal reset case the SM detector performs as well as the LRT one. Indeed SM is the optimal strategy for exponential distributions because the sample mean is a sufficient statistic [51], and the statistics $\Lambda(\underline{\tau})$ and $\mathscr{A}(\underline{\tau})$ coincide.

To simplify the performance analysis of the detector we introduce a synthetic index. To this aim, we consider the intersection between the ROC and the bisector of the first quadrant angle, which is very close to the point of ROC curve with the minimum distance from the axis origin. At this point $P_f = P_m$, and we can unambiguously define the error rate $P_e \equiv P_f = P_m$, which is representative of the detector behavior[1].

---

[1] In ref. [1] it has been shown that, for large sample size $N$, it is possible to relate $P_e$ with the well-known Kumar-Carrol index $d_{KC}$ [36].

To characterize the detectors, it is necessary to specify the time $T_{obs}$ in which the $N$ escape times are collected. The average escape time depends upon the hypothesis in force, and so does the number of escapes collected in a fixed time interval. However, to properly use the LRT strategy in its classical formulation the sample size has to be *fixed*. We have adopted the solution to compute $N$ with constant mean observation time under $H_0$ hypothesis, $E[T_{obs}|H_0]$. In the incoherent case with the instantaneous signal reset strategy, we have

$$N = \frac{E[T_{obs}|H_0]}{\mu_0}. \tag{12}$$

Equation (12) is to be changed for the coherent case because of the time lost to reset the signal with the same initial phase. This loss leads to the approximated relation [1]:

$$N \approx \begin{cases} \dfrac{E[T_{obs}|H_0]}{2\pi/\omega}, & \mu_0 \leq \dfrac{\pi}{\omega} \\ \dfrac{E[T_{obs}|H_0]}{\mu_0 + \pi/\omega}, & \mu_0 > \dfrac{\pi}{\omega} \end{cases} \tag{13}$$

In the incoherent case with random signal reset strategy, it is hard to map the mean observation time to a fixed number of escapes. To perform a fair comparison, we have supposed that no more of a signal period is lost. So the relationship is:

$$N = \frac{E[T_{obs}|H_0]}{\mu_0 + \pi/\omega}. \tag{14}$$

Another more challenging issue arises when using LRT that, as evident from the Eq. (11), improves detection by using the whole PDF. Unfortunately, the escape time distributions are not theoretically known for the system described by Eq. (4). Even in the case $S_0 = 0$ the Arrhenius law is approximately valid for rare escapes [17] (in the unperturbed oscillator time scale $\omega_J$), while for fast escapes (that are interesting for signal analysis) only approximated analytical estimates exist [53]. When the signal is applied the knowledge of the escape time distributions is even poorer, and essentially limited to the overdamped case [9]. Thus we are compelled to numerically estimate the PDFs; however, it is possible to do better than just accumulate data with extensive simulations using a suitable non-parametric statistical technique, such as the Kernel Density Estimation (KDE) [54]. This technique generalizes the basic idea of histogram by means of a so-called Kernel function $K(\cdot)$, usually a symmetric PDF (we employ a standard normal distribution). For the random sample $\underline{X} = \{X_i\}_{i \in [1,N]}$ of size $N$ the kernel estimator is

$$\widehat{g}(x) = \frac{1}{Nw} \sum_{i=1}^{N} K\left(\frac{x - X_i}{w}\right). \tag{15}$$

The parameter $w$ is the *bandwidth* (also called *smoothing parameter*). The optimal choice for this parameter depends on the sample size, the kernel and the PDF which has to be estimated. In several scenarios it is computed by means of the relation

$$w = \left(\frac{4}{3}\right)^{1/5} \widehat{\sigma} N^{-1/5}, \tag{16}$$

where $\widehat{\sigma}$ is the sample standard deviation of the random sample. In applying this framework to escape times, we immediately encounter a first difficulty. Escape times are *positive*, i.e. the PDFs, under both hypothesis $H_{0,1}$, are:

$$f_j(t) = 0, \quad \forall t < 0. \tag{17}$$

On the contrary, Eq. (15) leads to an estimated PDF which does not satisfy the inequality (17). To go around this problem we can transform escape times $\underline{\tau} = \{\tau_i\}_{i \in [1,N]}$ according to

$$X = \log(\tau). \tag{18}$$

The obtained random sample $\underline{X} = \{X_i\}_{i \in [1,N]}$ can assume every value on the real axis and the PDFs $\widehat{g}_j(x)$ can be estimated with Eq. (15). Finally, the PDF $\widehat{f}_j(t)$ is obtained from $\widehat{g}_j(x)$ via

$$\widehat{f}_j(t) = \frac{\widehat{g}_j(\log(t))}{t}, \quad t > 0. \tag{19}$$

The procedure is applied to both PDFs, $\widehat{f}_0(\cdot)$ and $\widehat{f}_1(\cdot)$, with ($H_1$) and without ($H_0$) the deterministic signal. The estimates are finally inserted in the statistic, Eq. (11):

$$\widehat{\Lambda}(\underline{\tau}) = \frac{1}{N}\sum_{i=1}^{N} \log\left[\frac{\widehat{g}_1(\log(\tau_i))}{\widehat{g}_0(\log(\tau_i))}\right] \begin{array}{c} H_1 \\ > \\ < \\ H_0 \end{array} \zeta, \tag{20}$$

To obtain reliable estimations a large sample size ($\sim 5 \times 10^5$) has been used. We observe a slight over-smoothing, whose effect is some worsening of performances for LRT. Indeed the PDF oscillations, which contain most of the additional information with respect to the sample mean, are underestimated. This approach is to be considered *conservative*, because the presence of artifacts in the opposite under-smoothing case could artificially improve the LRT performances, leading to over-estimated results.

## 4 Josephson Junction Detector: Performance Evaluation

The application of the detection theory of Sect. 3 to JJ allows to improve the possibility of exploiting JJ to reveal signals. The quality of the detection—the parameter $P_e$—can greatly vary with the same noise-signal combination (or SNR) applied to JJ with different parameters (dissipation, capacitance, constant bias). Let us summarize the parameter region in which to expect the best detection performances. In discussing the basic Eq. (2), we have emphasized that JJ are characterized by four electrical quantities: the bias current $I_b$, the Josephson critical current $I_c$, the junction resistance $R$ and capacitance $C$. The optimization of a JJ as a detector leads to the following results [1, 23]:

1. The dc current $I_b$ can be assumed positive (for the symmetry of the problem) and below the critical current [to have two solutions, see Eq. (6)]: $0 \leq I_b \leq I_c$. In normalized units the interval reads $0 \leq \gamma \leq 1$. The optimal bias current $I_b$ for the LRT strategy should be set as close as possible to $\gamma = I_b/I_c \simeq 1$ to achieve the lowest value of $\Delta U$. For the SM strategy the currents should be set at an intermediate value that depends on the ratio between the energy barrier and the signal.
2. The critical current $I_c$ normalizes (among the other quantities) the signal amplitude and noise, $\varepsilon$ and $\varepsilon_n$, respectively. This normalization, as expected, does not change the SNR which is connected to $\varepsilon/\sqrt{\varepsilon_n}$. However it is found that, at a given level of SNR, larger normalized noise intensity favors detection. The limit for the parameter $\varepsilon_n$ is to keep negligible the number of escapes of the system toward the higher local minima of the potential [see Eq. (5)], i.e. toward the stable points at the left in Fig. 1
3. The resistance $R$ normalizes the dissipation parameter $\alpha$. A low dissipation is beneficial because results in shorter average escape time, and therefore increases the statistic numerosity at a fixed length of the analyzed signal.
4. The capacitance $C$ determines the normalized frequency $\omega = \Omega/\omega_J$ of the signal. The best performances occur around the geometrical resonance of Eq. (4), $\omega \simeq \omega_{res}$, for both LRT and SM. JJ are therefore best suited for signals whose frequency is in the range $10 - 1000\, GHz$.

The last property, the behavior of the JJ detector as a function of external signal frequency deserves special attention. It amounts to use the JJ as a nonlinear filter, with a peculiar resonance figure. In Fig. 5 it is displayed the error rate $P_e$ for the coherent SM an LRT strategy, compared with the same error for the incoherent detection strategy in the instantaneous signal reset case.

Figure 5 clearly indicates that a minimum error probability occurs at the resonance frequency $(e)$ determined through linearization of Eq. (4) for small signal amplitude $\varepsilon$ [4]:

$$\omega_{res} \simeq \left(1 - \gamma^2\right)^{1/4}. \qquad (21)$$

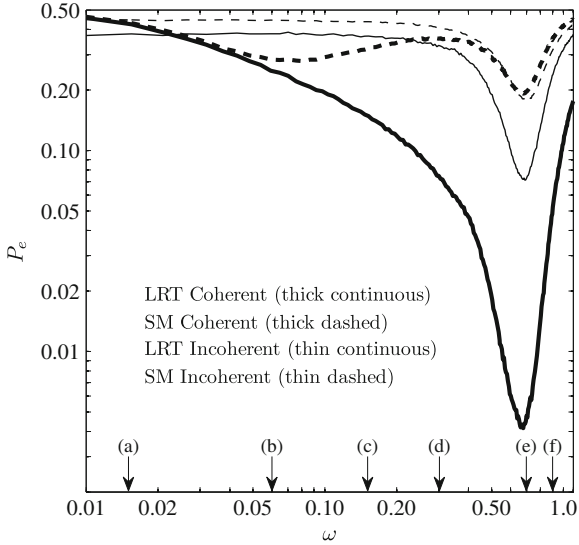

**Fig. 5** The error rate $P_e$ versus the angular velocity $\omega$ of the applied signal. The curves represent: LRT Coherent (thick continuous line); SM Coherent (thick dashed line); LRT Incoherent with instantaneous signal reset (thin continuous line); SM Incoherent with instantaneous signal reset (thin dashed line). The labeled arrows indicate the frequencies of the PDF shown in Fig. 6. Parameters of the simulations are: $\gamma = 0.8$, $\alpha = 0.05$, $\varepsilon_N = 0.0175$. The signal parameter are $\varepsilon = 0.05$ and $\varphi_0 = 0$. The mean observation time under $H_0$ is $E[T_{obs}|H_0] = 2000$

For non vanishing $\varepsilon$ nonlinear corrections to the second order Taylor expansion tune the resonance: $\omega_{res} = \omega_{res}(\varepsilon)$ [32]. The analysis of JJ detectors confirms the dependence predicted by Eq. (21); there exists a suitable neighborhood of $\omega_{res}$ that is one of the best regions for detection purpose, with a small correction for the finite signal amplitude. The phenomenon is associated with a multivalued resonance curve of the non linear plasma frequency [37]. It is remarkable, to our opinion, that best detection performances (in terms of error probability $P_e$) are reached in the same strongly non linear distortion regime as in amplifiers [58].

Focusing on the coherent case, we note that for SM strategy a second interesting region occurs at a lower normalized signal frequency $\omega_{SR}$ (see (*b*) in Fig. 5), where another resonance appears at a frequency that can be also much smaller than the geometric resonance. The position of this resonance dip depends upon the phase and signal temperature [23], so it can be considered a stochastic resonance [8, 26, 27, 44] or resonant activation [19, 21, 57]. In Ref. [23] it has been found that a region of optimal detection is pinpointed if the potential well barrier (tuned by $\gamma$), the normalized signal frequency $\omega$ and the normalized noise intensity $\varepsilon_n$ are connected by the relation ($C(\varphi_0)$ is a function of the initial phase $\varphi_0$ and $\mu_0$ is the average escape time under the hypothesis $H_0$):

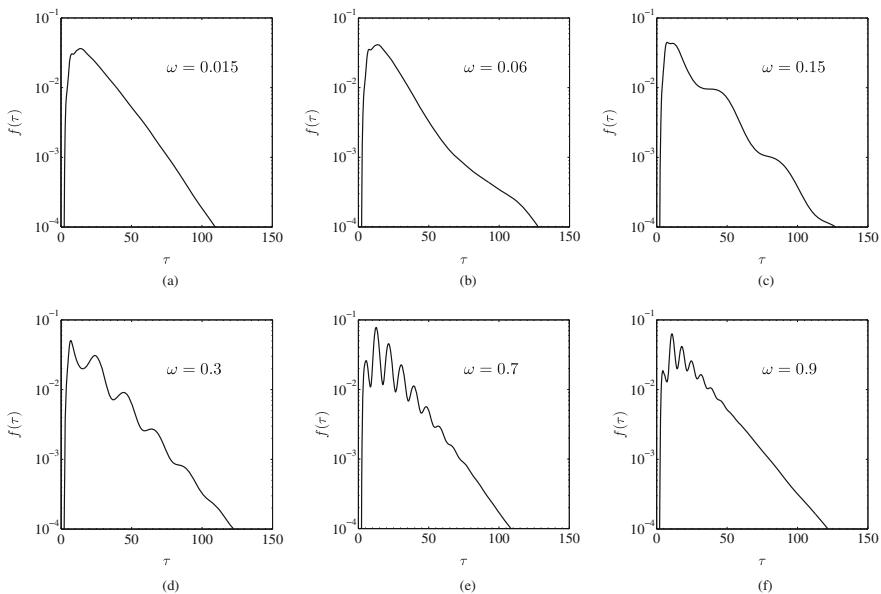

**Fig. 6** PDFs of the escape times acquired by means of the coherent strategy at different applied frequency (see the arrows at the bottom of Fig. 5). **a** $\omega = 0.015$. **b** $\omega = 0.06$ (stochastic resonance). **c** $\omega = 0.15$. **d** $\omega = 0.3$. **e** $\omega = 0.7$ (geometric resonance). **f** $\omega = 0.9$. Parameters of the simulations are the same as in Fig. 5

$$\omega_{SR} = \frac{\mu_0}{2\pi C(\varphi_0)} \exp\left\{\frac{2}{\varepsilon_n}\left[\sqrt{1-\gamma^2} - \gamma \cos^{-1}(\gamma)\right]\right\}. \tag{22}$$

Below the stochastic resonance frequency the PDFs computed under $H_1$ and $H_0$ are very similar, while above this frequency the PDFs computed under $H_1$ develop oscillations (see Fig. 6). This explains the disappearance of stochastic resonance of Eq. (22) in the coherent LRT detection framework, for LRT exploits the PDFs oscillations and does not deteriorate above the frequency (22). To see this, let us imagine to increase the applied frequency $\omega$ around $\omega_{SR}$, marked by the label (*b*) in Fig. 5. The analysis of the PDF of Fig. 6 reveals that the curves shown in Fig. 5 have an intuitive explanation. In fact the crucial passages across the SR (Fig. 6a,b,c) and the geometric resonance (Fig. 6d,e,f) are marked by a qualitative change of the PDF. At the SR frequency $\omega_{SR}$ (Fig. 6b) the slope (and hence the average escape time) is most changed, while above $\omega_{SR}$ (Fig. 6b) the oscillations are clearly visible. This is the reason why, in the coherent case, SM and LRT methods have so different behaviors. Indeed coherent SM, that just detects the average, is most sensible to the changes of the slope and exhibits a minimum (i.e. best performances) in correspondence of this point. Conversely LRT, that takes into account "all" details of the PDF's, can improve even when the first momentum ceases to be an effective statistics for hypothesis discrimination. For completeness, let us point out that inspection of Eq. (22) reveals that $\omega_{SR}$ decreases when the normalized noise intensity $\varepsilon_n$ increases, and therefore a fixed value of the external applied frequency has the effect to increase the efficiency of the SM. Such paradoxical increase of the performances increasing the noise level for the SM is solved by the observation that the improvement obtained at the $\omega_{SR}$ frequency can be outperformed by the choice of a more refined detection strategy (the LRT) that takes into account the PDFs oscillations. In this sense it is not really possible to improve detectors increasing the noise. Indeed, see Fig. 5, the coherent SM detector performances are always worse than the coherent LRT performances, and this also confirms the general idea that stochastic resonance is a consequence of a suboptimal detection scheme [25]. The practical consequence is that synergetic effects leading to stochastic resonance between noise and signal in JJ devices can only be exploited in suboptimal strategies such as SM, while in optimal detection strategies noise should only be reduced as much as the experimental set up allows.

Turning to the incoherent detection strategies, it is evident that their performances are worse than the corresponding coherent ones [23]. In any case, due to the residual oscillation of the PDF under $H_1$ hypothesis, the LRT approach guarantees better performances than the SM for the instantaneous signal reset case. Moreover, it is remarkable that the stochastic activation phenomenon disappears for SM, as a consequence of the reshuffling of the initial phase in the reset procedure of the JJ.

Performances for the random signal reset case are not shown in Fig. 5. However the SM incoherent curve represents quite well the behavior of both SM and LRT. As expected, LRT does not significantly improve the SM, due to the absence of oscillations.

In the above performed analysis we have fixed both the average time observation window $E[T_{obs}|H_0]$ and the SNR (related to the ratio $\varepsilon/\sqrt{\varepsilon_n}$). On the other hand, they strongly affect the error rate; for large sample size $N$ the error rate can be expressed as [1]:

$$P_e = \frac{1}{2}\text{erfc}\left[B_Y\sqrt{E[T_{obs}|H_0]}\left(\frac{\varepsilon}{\sqrt{\varepsilon_n}}\right)^{\eta_Y}\right] \quad (23)$$

where $B_Y$ and $\eta_Y$ depend on the particular detection strategy ($Y = \mathscr{A}, \Lambda$, for the SM and LRT respectively). If we focus on the scaling law index $\eta_Y$, it is remarkable that, for the coherent LRT case, it exhibits the *nearly optimal* behaviour $\eta_\Lambda \approx 1$. Indeed this is also the behaviour of the *ideal* matched filter that directly analyzes the signal (and not the escape times) and that attains the optimal detection performance [33]. The proposed LRT coherent strategy never outperforms the matched filter, because the value of $B_\Lambda$ is not optimal [1]. The performances of SM strategy are poorer, especially for small signals, because $\eta_\mathscr{A} \approx 3/2$, as shown in [23]. This type of analysis can be reformulated in more intuitive fashion as follows. By lowering the ratio $\varepsilon/\sqrt{\varepsilon_n}$, the average observation time window $E[T_{obs}|H_0]$ should be increased to preserve the same error rate according to the law

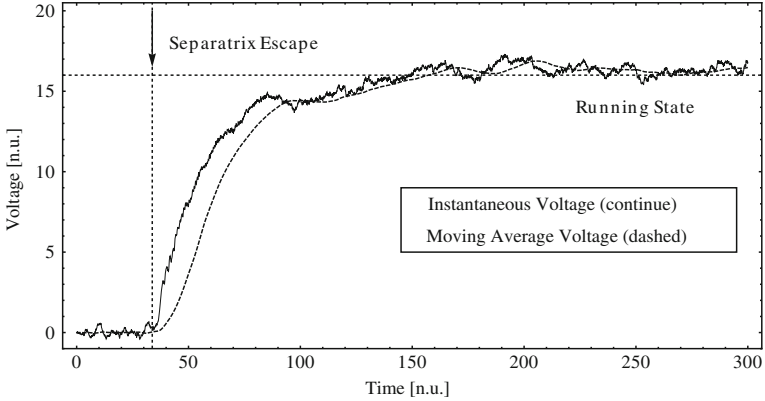

**Fig. 7** Time dependence of the instantaneous and moving average voltage. The average is computed over two periods of the Josephson frequency, see Eq. (25)

$$E[T_{obs}|H_0] \propto \left(\frac{\varepsilon}{\sqrt{\varepsilon_n}}\right)^{-2\eta_Y} \qquad (24)$$

Therefore for low SNR, the SM detector (characterized by $\eta_\mathscr{A} = 3/2$) is very inefficient with respect to the LRT detector (characterized by $\eta_\Lambda = 1$) because it requires a much longer observation time to achieve the same error rate.

## 5 Practical Issues for Josephson Junction Detectors

We have so far described the analysis of the escape times across the separatrix of Fig. 1. This corresponds to set a threshold of the phase at a fixed level, see Fig. 2, roughly related to the appearance of a finite voltage, Eq. (8). To transform such idea into practical realization there are several limits that further deteriorate the performances. First, the passage across the separatrix does not cause immediately the appearance of nonzero average voltage, for real voltmeter will measure the voltage average over a finite time.

In Fig. 7 we compare the instantaneous voltage (continuous line) to the voltage computed with a moving average voltage over 2 periods of the Josephson frequency (dashed line). The averaging roughly corresponds to a passing bandwidth 2 times lower than the Josephson frequency $\omega_J$. In Fig. 7 the dashed vertical line indicates to the time when the phase crosses the separatrix and the instantaneous voltage starts increasing. The actually accessible voltage is the moving average of the instantaneous voltage of the Josephson Eq. (2):

$$<V> = \frac{V_0}{\Delta T}\int_{\tau-\Delta T}^{\tau}\left(\frac{d\varphi(\tau')}{d\tau'}\right)d\tau'. \qquad (25)$$

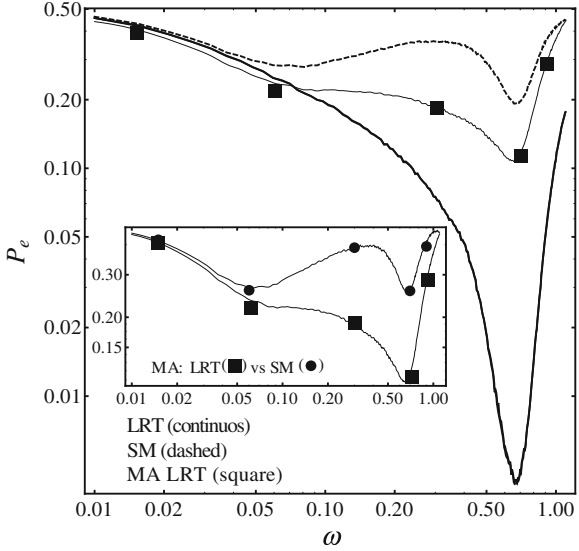

**Fig. 8** Error rate $P_e$ as a function of the applied signal frequency $\omega$. The line with square markers is computed for the LRT coherent strategy when the escape is defined through the moving average of the voltage, Eq. (25). For sake of comparison, the LRT (continuous line) and SM (dashed line) error rates, computed in the coherent case when the escape is defined through the passage across the separatrix, are also shown. In the inset it is shown a closeup of the LRT coherent (line with square markers) and SM coherent (line with circle markers), both computed when the escape is defined through the moving average of the voltage. Parameters of the simulations are: $\gamma = 0.8$, $\alpha = 0.05$, $\varepsilon_N = 0.0175$. Moreover, when the signal is present, $\varepsilon = 0.05$ and $\varphi_0 = 0$. The simulations are performed setting the mean observation time under $H_0$, $E[T_{obs}|H_0] = 2000$

The detection of an escape time is established with a suitable threshold on $<V>$. A realistic averaging is particularly important when the junction is in the so called intermediate damping regime ($0.25 < \alpha < 1$) and retrapping, that lowers $<V>$, occurs [42].

The smoothing and the consequent loss of information of the device are shown in Fig. 8 by a dashed line with square symbols. It is evident that LRT methods are also in this case better than the SM approach. Moreover, it is confirmed that stochastic resonance does not occur for optimized methods.

It is experimentally convenient to slowly ramp the bias current up to the critical value $I_c$ and to record the current at which the voltage appears. The escape time distributions $f_{0,1}(\tau)$ are related to the switching current distributions $P_{0,1}(\gamma)$. In the adiabatic approximation the relationship reads [24]:

$$P_{0,1}(\gamma) = \frac{1}{f_{0,1}(\tau)} \left(\frac{d\gamma}{d\tau}\right)^{-1} \left(1 - \int_0^\gamma P_{0,1}(\gamma')d\gamma'\right).$$

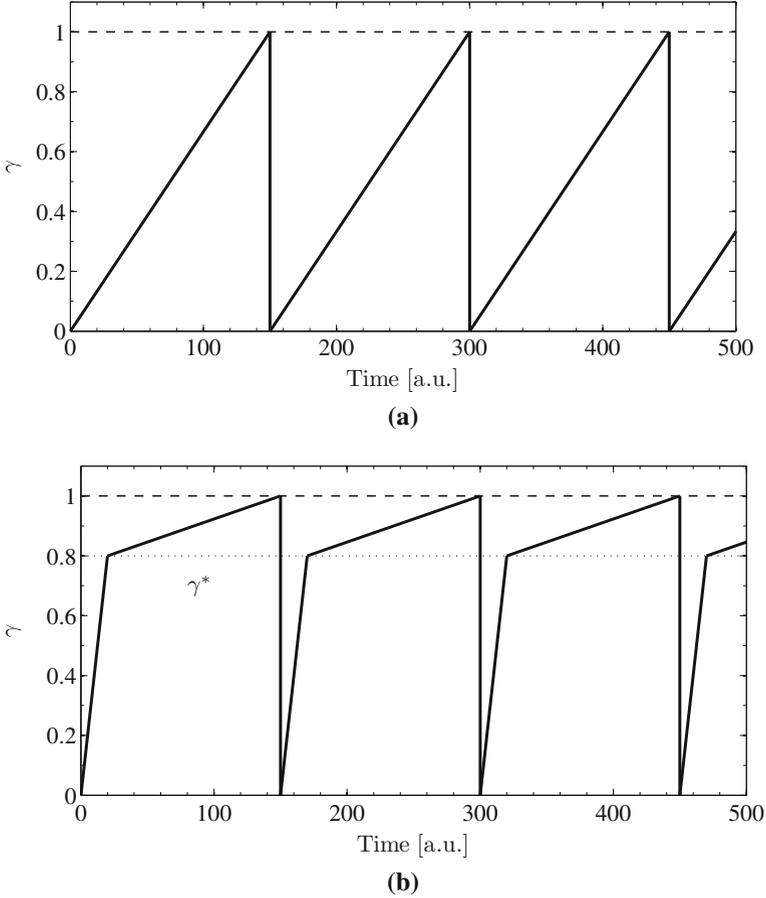

**Fig. 9** Top plot (**a**): current sawtooth waveform (continuous line) of the current $\gamma$. Bottom plot (**b**): current waveform (continuous line) with the maximum normalized current $\gamma$ (dashed line) and the starting point current level $\gamma^* = I_b^*/I_c$ (dotted line)

The advantage for the experiments is twofold. First, it is possible to use the ramping time as a sort of clock, and this enables an accurate timing easily available in the labs. Second, and more important, the sweeping bias method guarantees that a switch occurs during each ramp, thus fixing the time to collect the prescribed number of escapes.

In Fig. 9a the time dependent $I_b$ follows a simple sawtooth up to the critical current. When the switching current is close to the critical current the sawtooth bias method has the disadvantage that leaves the system for most of the time in the zero voltage state, and only collects escape times in a narrow region close to $I_c$. To avoid the depletion of so much time, a more sophisticated method could be employed, applying a profile of the type displayed in Fig. 9. In this case the bias current is quickly increased in the region where escape is not likely to occur, marked in Fig. 9 as $\gamma^* = I_b^*/I_c$. At this point the current sweep is slowed down to carefully examine the region $I_b^* < I_b < I_c$. Indeed in the second region the switching effectively occurs, and the data are consequently recorded. Our analysis roughly corresponds to reach very quickly (in a negligible time) $\gamma^*$ and to sweep much more slowly in the second region. However, apart some indications [30], a careful analysis of the consequences of the bias sweep on JJ detectors is at the moment still lacking. We remark that constant bias with an external clock allows time resolution down to nanoseconds and the range spans over six orders of magnitude [55].

Another relevant issue is related to the impossibility of perfect knowledge of the JJ and signal parameters. It is unrealistic to suppose that the initial phase $\varphi_0$ of the signal is known, while it is possible to exploit some guess, related to the particular application, about the signal amplitude $\varepsilon$ and frequency $\omega$. A viable solution, briefly analyzed in [1], is to use a Generalized LRT detection strategy [33] that performs a joint near—optimal signal detection and phase estimation. Unfortunately, due to the difficulties encountered in realizing perfectly reproducible JJ, also some physical parameters, such as the critical current $I_c$ or the capacitance $C$, are not perfectly known. We guess that a careful analysis of the escape time could be useful to individuate some phenomena, characterized by a well known *signature*, that can be used to reduce the uncertainty.

We conclude the Section with the role of the noise sources in JJ. The distribution of the escape times has been used to reveal the non-Gaussian character of the noise due to the granularity of weak bias currents [47]. In Eq. (2) quantum noise has been neglected. To include quantum fluctuations more than a simple differential equation is required. However, quantum fluctuations can be approximated by a correlated Gaussian noise, although there is not a single approach to the solution of the full model (as shown for instance in Refs. [12, 38, 52]). The noise spectrum is therefore an hallmark of quantum behavior; we speculate that the analysis of the escape times could be employed to discriminate if the noise is correlated, and thus to go beyond the use of the simple mean escape time. Moreover, it has been argued that several phenomena that are characteristic of quantum behavior could be reproduced by a classical model such as Eq. (2) through an appropriate choice of the temperature [14, 30, 31, 11]. We suggest that even in the case when the temperature of the junction is not certainly known, for example if the thermal contact with the bath is not perfect, the analysis of the full PDF with the methods of signal theory, rather than just compare the average escape time [11], could give indications for the correct model.

## 6 Conclusions

We have proposed to examine the escape time dynamics of JJs from the perspective of detection. Indeed from this standpoint JJs are potentially interesting for two main reasons: speed and low intrinsic noise (temperature). To assess the effectiveness of the detector, we have chosen the error rate as performance indicator of the capability of discerning between two different conditions (e.g. presence and absence of a sinusoidal excitation). Moreover, the real experimental set-up should be designed in view of the detection, calibrating the parameter of the junction ($I_c$, $C$, and $R$) and the external bias supply ($I_b$). Finally, the analysis of the data should be optimized to extract as much information as possible from the available measurements. In this framework it is possible to find several interesting features, for instance that low dissipation junctions are the most suitable for signal analysis, or that the bias point should be chosen to have such a low energy barrier as it is possible, compatibly with the speed of the electronics to keep trace of the escape. Also the feasible criterion to decide if an escape has actually occurred is important, because practical methods to detect a voltage change entail a loss of information that should be carefully controlled. As a concluding remark, we want to underline that signal detection is important for practical purposes as the identification of a waveform embedded in a noisy background, but the underlying theory, i.e. the statistical hypothesis testing, can also be used as a tool to distinguish between two models. Two noticeable examples are the retrapping regime for intermediate damping and the identification of the escapes due to quantum tunneling. In these circumstances, the detection theory approach allows to search for the most efficient way to distinguish which model is best corroborated by the data. The gain with respect to an analysis confined to intuitive statistics, such as the sample mean, could be considerable.

**Acknowledgments** We acknowledge Giacomo Rotoli for fruitful advices. We also thank Luca Galletti, Davide Massarotti and Francesco Tafuri for useful discussions. This work has been supported by the Italian Super Computing Resource Allocation ISCRA, CINECA, Italy (Grant IscrB_NDJJBS 2011).